\newcommand {\bea}{\begin{eqnarray}}
\newcommand {\eea}{\end{eqnarray}}
\newcommand {\be}{\begin{equation}}
\newcommand {\ee}{\end{equation}}

\documentstyle[12pt]{article}

\begin{document}
{\hbox to\hsize{ \hfill UPR-735-T}\par
{\hbox to\hsize{ \hfill hep-th/9702153 }\par
\begin{center}
{\LARGE \bf
A String Model of \\[0.1in]
Black Hole Microstates} \\[0.3in]
{\bf Finn Larsen}\footnote{e-mail: larsen@cvetic.hep.upenn.edu}\\[.05in]
{\it Department of Physics and Astronomy \\
University of Pennsylvania \\
Philadelphia, PA 19104} 
\end{center}

\begin{abstract} 
The statistical mechanics of black holes arbitrarily far from extremality
is modeled by a gas of weakly interacting strings. 
As an effective low energy description of black holes the string model 
provides several highly non-trivial consistency checks and 
predictions. Speculations on a fundamental origin of the model 
suggest surprising simplifications in non-perturbative string 
theory, even in the absence of supersymmetry.

\end{abstract}


Black holes exhibit thermodynamic properties suggestive of 
a complicated internal structure. This is somewhat paradoxical
within standard general relativity where black holes are absolutely 
featureless. In the last year or so there has been considerable 
excitement as it has become clear that string theory can accurately 
account for the degeneracy implicit in the Bekenstein-Hawking entropy.
This development initially concerned the counting of states for 
extremal~\cite{sen95,structure,strom96a} and near extremal black 
holes~\cite{callan96a,strom96b}. The effective string model that 
emerged apparently describes in much detail the dynamics of near 
extremal black holes~\cite{mathur,greybody,backreact,cgkt}. 
This no longer leaves any doubt that string theory describes many 
features of black holes.

Many questions remain however. Current models are restricted to 
very large wave lengths. In the interesting region with possible 
information loss the appropriate methods are quite different and 
they still need further development~\cite{cfthair1,hair1,cfthair3,hair2}.
The focus of this Letter is another ill-understood issue,
the generalization far off extremality. Here many new concerns appear, 
especially because much standard string technology remains 
untested in a regime that is not even approximately supersymmetric.
Some proposals have appeared (including~\cite{strom96d,susskind96,polch96} ) 
but there is still much confusion. Moreover, a large region of
parameter space has been identified where the $D$-brane motivated 
model gives incorrect predictions~\cite{hawking97,dowker}. 
In this Letter we shall 
argue that, nevertheless, the situation is quite hopeful: the 
thermodynamics of black holes in string theory has non-trivial 
features suggesting a surprisingly simple improved effective string model
that gives a satisfying description of non-extremal black holes. 
We shall present arguments motivating the string model and note
a number of consistency checks, including highly non-trivial ones.
We speculate that the model arises in fundamental string theory
as the direct product of two chiral sectors, each a $c=6$ 
superconformal field theory.
 
The discussion will assume five-dimensional black holes although
the generalization to four dimensions involves no new features. 
For definiteness we will presume toroidally compactified type II theory. 
In this context the most general non-rotating 
black hole is characterized up to duality by its 
mass $M$ and three conserved $U(1)$ charges $Q_i~;~i=1,2,3$~\cite{hull96}. 
It is convenient to introduce in intermediate steps the 
so-called non-extremality parameter $\mu$ and also the ``boosts'' 
$\delta_i~;~i=1,2,3$. In the resulting parametric form the physically 
interesting quantities 
are~\cite{cygeneral}\footnote{We employ units where $G_N={\pi\over 4}$.}
\bea
M&=&{1\over 2}\mu\sum_i \cosh 2\delta_i 
\label{eqn:mass} \\
Q_i &=&{1\over 2}\mu \sinh 2\delta_i 
\label{eqn:charge}\\
S&=& 2\pi\mu^{3\over 2}\prod_i \cosh \delta_i~.
\label{eqn:entropy}
\eea
When the black hole solution is embedded in string theory the 
charges arise from specific sources.
For example they may be the winding and momentum charges of a 
fundamental string as well as solitonic $5$-brane charge. 
Or they may be 1-brane and 5-brane RR-charges along with 
Kaluza-Klein momentum. The latter representation is particularly popular 
because the sources of RR-charges are $D$-branes which are accessible 
to detailed study in string theory~\cite{polch95a}. 
For the present purposes it will not be necessary to make a specific 
choice between these embeddings ( or many others).
In fact, non-perturbative string theory in the absence of supersymmetry 
remains largely unexplored and offers few concrete tools at this point. 
We will simply rely on the conjectured duality symmetries to ensure that 
microscopic  degeneracies will be identical in all representations. 
More importantly, duality also gives some guidance in the arguments. 
Note that the manifestly symmetric appearance of the three charges in the 
thermodynamic formulae eq.~\ref{eqn:mass}-\ref{eqn:entropy} is a 
necessary condition for this strategy to be consistent.

Physical charges $Q_i$ occur only in quantized units. The constants 
of proportionality relating $Q_i$ to integers $n_i$ depend on the precise 
embedding in string theory. Indeed fundamental charges, 
RR-charges, and solitonic
charges depend differently on the microscopic coupling constant, and
the geometric moduli of the compactified torus also enter. 
It is remarkable however that, for the black holes embodied
in eqs.~\ref{eqn:mass}-\ref{eqn:entropy}, the moduli
cancel in the product of charges so
\footnote{We have chosen normalizations to avoid a numerical 
factor in this formula.} 
\be
\prod_i Q_i = \prod_i n_i~.
\label{eqn:Qin}
\ee
Generically physical parameters transform
in covariant but nevertheless complicated ways under the duality symmetry.
The right hand side of eq.~\ref{eqn:Qin} is very special in this
respect because here the full duality symmetry group could 
easily be restored. Indeed, $\prod_i n_i$ is the manifestation, for
our representative class of solutions, of the unique cubic 
invariant of the $U$-duality group $E_{6(6)}$~\cite{dvv1}. 
The cancellation of moduli responsible for this simplification 
is of great importance because it suggests a direct relation between 
macroscopic quantities and the underlying microscopic 
theory~\cite{structure,strom96c,ferrara}. This is particularly clear 
in the extremal limit $\mu\rightarrow 0$, $\delta\rightarrow\infty$ 
with $Q_i$ fixed. Here the entropy becomes  
\be
S=2\pi (\prod_i Q_i)^{1\over 2}=2\pi (\prod_i n_i)^{1\over 2}~.
\ee
Note that this is also the degeneracy of a chiral string model with
central charge $c=6$ and an effective level $N=\prod_i n_i$. 
For supersymmetric black holes such a model is very well motivated from
fundamental string theory. 

In the general non-extremal case we do not {\it a priori} know the
moduli dependence of the mass eq.~\ref{eqn:mass} and the entropy
eq.~\ref{eqn:entropy}. It is possible that the microscopic quantum
theory ensures that even among these macroscopic parameters there are
moduli independent combinations, interpretable as quantum numbers. 
We propose that indeed there is exactly {\it one} such quantity and 
it is the entropy. Off course
it is the exponential of the entropy that might actually be an integer. 
The point is however that, by definition, quantum numbers do not depend 
on moduli. It is a consequence of the proposal that, as moduli vary 
with quantum numbers fixed, the mass changes in a rather complicated way, 
as implied by eqs.~\ref{eqn:mass}--\ref{eqn:entropy}.
The idea that mass depends on moduli but degeneracies do not is familiar
from perturbative string theory where degeneracies follow from central 
charge and level of the conformal field theory, whereas mass depends on 
both level and (moduli dependent) physical charges. 
Our working assumption can be motivated as follows: the semiclassical
derivation of the entropy formula shows a remarkable independence
of details, such as the low-energy matter content. We interpret this 
universality as a macroscopic manifestation of the underlying  
quantum theory. Formally, we note that quantization conditions 
on the gauge charges derive from topologically non-trivial 
gauge transformations 
at infinity~\cite{hair2}; so we expect an analogous condition to 
follow from general covariance. In fact, the Gibbons-Hawking derivation of 
the entropy formula from the functional integral appears to realize
exactly this~\cite{hawkingtwo}. 

At this point it is instructive to consider briefly the more general 
case of rotating black holes. In five dimensions the rotation group is 
$SO(4)\simeq SU(2)_R\times SU(2)_L$~; so configurations are characterized
by two angular momenta $J_{R,L}$. The entropy is~\cite{rotation3}
\be
S=2\pi(\sqrt{N_R}+\sqrt{N_L})
\label{eqn:snrnl}
\ee
where
\be
N_{R,L}= 
{1\over 4}
\mu^3(\prod_i\cosh\delta_i\mp\prod_i\sinh\delta_i)^2-J_{R,L}^2~.
\label{eqn:nrl}
\ee
In string theory the space of states is the direct product of two
terms because of the world sheet decomposition into left and right 
moving modes. Entropy deriving from string degeneracy is 
therefore expected to be the sum of two terms. It is suggestive
that eq.~\ref{eqn:snrnl} takes this characteristic form. 
Concretely we propose that, in analogy with the supersymmetric case,
$N_{R,L}$ arise as effective levels of two independent chiral sectors 
of a single string theory. Assuming central charge $c=6$ in
each sector eq.~\ref{eqn:snrnl} expresses a quantitative microscopic 
interpretation of the entropy in the most general case. 
It generalizes in a duality invariant fashion the $D$-brane motivated
effective string model (references include~\cite{callan96a,mathur97}).
A consistency requirement on this proposal is that $N_{R,L}$ 
are separately quantized with all positive integer values as 
spectrum. Since the angular momentum is quantized it would certainly
take a very complicated conspiracy to make the total entropy 
moduli independent without $N_R$ and $N_L$ being so separately. 
Moreover, the angular momenta were normalized in the conventional way 
taking all integer values; so if indeed the first term in 
eq.~\ref{eqn:nrl} is quantized with all integers as spectrum,
the full $N_{R,L}$ will also be so for all values of the
angular momenta. This is an important test of the proposal.
It also implies that, accepting the proposal in the absence of
angular momenta, generalization to rotating black holes 
is automatic, with numerical factors matching 
correctly~\cite{rotation1,rotation2}. 
Vanishing angular momenta can therefore be assumed in the following, 
without loss of generality. Angular momenta nevertheless play an 
important auxiliary role in the argument because they indicate how 
the entropy eq.~\ref{eqn:entropy} should be divided into left and 
right moving contributions, and also because they suggest the precise 
quantization condition. The first point was noted and repeatedly emphasized 
by Cveti\v{c}, most recently in~\cite{cveticreview}.

It was argued from general relativity that, apart from charges,
there should be only one quantum number visible in the
macroscopic theory, namely the entropy. On the other hand our
concrete string model introduces two quantum numbers, the levels of 
the left and right moving states. The reason that there is no conflict 
is the constraint
\be
N_L - N_R = \mu^3 \prod_i \cosh\delta_i \sinh\delta_i 
= \prod_i Q_i = \prod_i n_i
\label{eqn:matching}
\ee
that follows from, in turn, eqs.~\ref{eqn:nrl},\ref{eqn:charge},
and~\ref{eqn:Qin}. This not only avoids an apparent contradiction 
but it also relates 
quantities, purely within the microscopic theory, in a fashion 
reminiscent of the matching condition in perturbative string theory. 
It is indeed natural to suspect that in the full non-perturbative 
theory some condition arises from world sheet reparametrization invariance. 
Duality is a powerful restriction on its possible form
and~eq. \ref{eqn:matching} is the simplest expression consistent
with duality. As an independent consistency check note that
both sides of eq.~\ref{eqn:matching} have all 
integers as spectrum. These non-trivial and desired results 
lend some support to the underlying assumptions.

In the preceding it was argued that eq.~\ref{eqn:entropy} should be
represented as the sum of two terms
\be
S_{R,L}=\pi\mu^{3\over 2}
(\prod_i\cosh\delta_i\mp\prod_i\sinh\delta_i)
\ee
that are separately quantized. The difference of the two 
terms is
\be
S_{-}= 2\pi\mu^{3\over 2}\prod_i \sinh\delta_i~.
\ee
It is amusing to note that this is ${1\over 4}$ the area of the inner
horizon. Hence, whereas the Bekenstein-Hawking formula suggests that 
somehow the {\it outer} horizon is quantized in Planck units the
present proposal implies that also the {\it inner} horizon is quantized. 
This is a tantalizing prospect, although the precise geometric 
significance escapes us at this point.

In the string model there are separate left and right moving gases of weakly 
interacting strings. The terminology of left and right moving 
modes is independent of whether there is a net Kaluza-Klein momentum: 
all charges are treated democratically.
Independent inverse temperatures of left and right moving modes follow
from the thermodynamic relations
\be
\beta_{R,L}=
({\partial S_{R,L}\over\partial M})_{Q_i}=
2\pi\mu^{1\over 2} (\prod_i\cosh\delta_i\pm\prod_i\sinh\delta_i)
\label{eqn:betarl}
\ee
with the physical (Hawking) temperature given by 
$\beta={1\over 2}(\beta_R+\beta_L)$. 
Assuming weak coupling between left and 
right sectors the emission 
spectrum is expected to be proportional to the product of the left 
and right occupation numbers and, specifically, their characteristic 
statistical occupation factors. For this to be possible while, 
at the same time, the black hole exhibit the correct overall 
thermality predicted by Hawking, 
certain grey body factors must take a very special form. The 
absorption cross-section of a black hole is calculable 
in classical field theory and the grey body factor is easily
extracted. Remarkably, Maldacena and Strominger found a regime
where it has exactly the right form for this 
interpretation to be consistent~\cite{greybody}. For our purposes it 
is the precise value of the temperatures that is of interest as they
lead to a test of our ideas. The temperatures are only known in the
near extremal limit $\delta_i\gg 1$ where~\cite{hawking97,dowker}
\be
\beta_{R,L}\simeq 2\pi\mu^{1\over 2} 
[(\prod_i\sinh^2\delta_i+
\sum_{i<j}\sinh^2\delta_i\sinh^2\delta_j)^{1\over 2} 
\pm\prod_i\sinh\delta_i]~.
\label{eqn:greyrl}
\ee
In its region of validity
eq.~\ref{eqn:greyrl} agrees with the prediction eq.~\ref{eqn:betarl}.
This is particularly satisfying because the $D$-brane model only 
accounts for the temperatures when, in addition to $\delta_i\gg 1$, one 
boost parameter is much smaller than the other two, say
$\delta_3\ll \delta_{1,2}$~\cite{hawking97,dowker}. Moreover, the 
model here provides a precise prediction for the general case:
when $\delta_i \sim 1$ the expression under the square root should be
augmented with terms that makes it a perfect 
square $\prod_i(\sinh^2\delta_i+1)=\prod_i \cosh^2\delta_i$. 

The black hole also interacts with its surroundings through absorption
and emission of charged particles. We consider a particle 
of the species referred to by index $3$. Statistical properties 
of charged particles are more complicated
because they depend on both the energy $\omega$ and the charge $q_3$ 
of the particle. The combined effect is captured by statistical
distributions with characteristic exponent 
$\beta(\omega-q_3\Phi_3)$. Potentials follow from 
eqs.~\ref{eqn:mass}-\ref{eqn:entropy} and are given 
by
\be
\Phi_i=
-{1\over\beta}({\partial S\over \partial Q_i})_{Q_{j\neq i},M}
=\tanh\delta_i~.
\label{eqn:phii}
\ee
In the string model the background charge $Q_3$ couples to both left 
and right sectors; so the effective left and right moving 
distribution functions have exponents that exhibit intricate structure. 
Using the left and right entropies, respectively, and proceeding as in
eq.~\ref{eqn:phii} we find
\be
[\beta(\omega-q_3\Phi_3)]_{R,L}=\pi\mu^{1\over 2}
[(\omega-q_3\tanh\delta_3 )\prod_i \cosh\delta_i 
\pm
(\omega-q_3\coth\delta_3 )\prod_i \sinh\delta_i ]~.
\label{eqn:chargeexp}
\ee
Analogous exponents were inferred from the absorption cross-section 
in the near extremal limit~\cite{dowker}. They are
\bea
&~&[\beta(\omega-q_3\Phi_3)]_{R,L} \simeq 
\pi\mu^{1\over 2}\prod_i\sinh\delta_i \times 
\label{eqn:chargeex}\\
 &\times& [\left( 
(\omega-q_3\coth\delta_3)^2 
(1+{1\over\sinh^2\delta_1}+{1\over\sinh^2\delta_2})
+{\omega^2-q^2_3\over\sinh^2\delta_3}
\right)^{1\over 2} 
\pm(\omega-q_3\coth\delta_3) ]~. \nonumber
\eea
It is straightforward to check that only terms of
order at most $\sinh^{-4}\delta_i$ for large $\delta_i$ need
to be added under the square root in order that the expression
becomes a perfect square, the square root can be taken, and 
eq.~\ref{eqn:chargeexp} follows. Hence eq.~\ref{eqn:chargeexp} 
agrees with eq.~\ref{eqn:chargeex} in the regime where 
eq.~\ref{eqn:chargeex} is valid. This detailed agreement
between complicated functions is a highly non-trivial check 
on the string model. The unusual situation where the approximate 
expression is much more complicated than the proposed exact one 
suggests some unrecognized underlying simplicity.

The mass is not an independent parameter, but rather a definite 
function of the charges and levels given implicitly by 
eqs.~\ref{eqn:mass}-\ref{eqn:entropy}. In a complete microscopic 
theory this relation should be derived from arguments internal to 
the theory. It seems plausible that this should be possible, using 
dualities and Lorentz invariance. Incidentally, it should
be noted that the arguments in the preceding paragraphs relied on 
no ``mass renormalizations'' or ``redshifts'' and indeed none are 
allowed by the agreement with grey body factors. Therefore
it can be argued in the converse direction that the string model
parametrize the low energy interaction of the non-extremal black 
holes.

Up to this point the string model has been presented as an effective 
theory that embodies the statistical mechanics of non-extremal black 
holes. It should correctly capture the low energy 
physics of black holes in string theory. Although the
evidence presented in its favor is quite strong the precise relation to
more conventional ideas in string theory is certainly not clear.
To elucidate the question consider a bound state of $n_1$ 
$D1$-branes and $n_5$ $D5$-branes, 
carrying momentum $k$. Strominger and Vafa~\cite{strom96a} proposed 
that, when wrapped on $K3\times S_1$, the exact degeneracy of 
this supersymmetric state is described by level $k$ of a superconformal 
sigma-model with target space
\be
C = (K3)^{\otimes(n_1 n_5+1)}/\Sigma(n_1 n_5+1)
\ee
where the quotient by the symmetric group is implemented by
orbifolding the product manifold. 
An unusual feature is that, rather than being a fixed numerical 
constant, the central charge depends on the quantum numbers of
the configuration at hand. The formula is not manifestly duality invariant, 
as the three charges do not appear symmetrically. An alternative approach,
due to Dijkgraaf, Verlinde, and Verlinde~\cite{dvv1,dvv3}, 
was motivated by ideas about second quantized string theory and 
yields manifestly duality invariant expressions. It was subsequently 
shown that the proposed degeneracies are in fact identical~\cite{dmvv}
and specifically the Strominger-Vafa formula is consistent with duality.
The rather involved mathematics describing the space of {BPS} states 
in various cases is currently being explored~\cite{harveymoore}.
For our purposes we note that, for large charges, the degeneracies 
derive from a superconformal field theory with a more familiar, 
state independent, central charge $c=6$, but an unusual relation 
between the level and spacetime charges, namely $N=\prod_i n_i$. 
As explained earlier the corresponding string degeneracy 
accounts for the entropy of extremal black holes. The point here is 
the following: statistical mechanics of non-extremal black holes 
suggests that the generalization to the non-supersymmetric context 
simply involves two such structures with the full space of states 
realized as the direct product. Each sector describe $BPS$ excitations
of some brane configuration wrapped on compact manifolds, not
necessarily the same on the two sides. Moreover, each sector is 
supersymmetric and duality invariant (or covariant, according to the 
precise context) and the two sectors are related by a matching condition 
eq.~\ref{eqn:matching} (suitably generalized) that respects these 
features. As usual the condition on the $0$-modes that relates this 
microscopic physics to a specific spacetime interpretation introduces moduli. 
Generically duality transformations do not leave moduli invariant; 
so the duality symmetry is spontaneously broken. 
In fact it is broken to a compact subgroup and this is 
quite welcome, as explicitly realized non-compact symmetry is unstable. 
The novelty in the present proposal is that the introduction of moduli also 
spontaneously breaks supersymmetry. As the analogous breaking of
duality is no cause of concern, it is reasonable to expect that 
also the breaking of supersymmetry is sufficiently mild 
that string theory remains under full control. 

It should be cautioned that the scenario in the previous paragraph
has only been realized in the regime where quantum numbers are large
and black hole physics is a guide. Clearly much needs to be done 
to make the proposal precise in the context of the full algebra of $BPS$ 
states and to establish the consistency of the resulting vacua
in the interacting theory. However, if this endeavor should be 
successful it could be quite rewarding: it would 
provide a precise tool for non-perturbative string theory in the 
absence of supersymmetry.

\vspace{0.2in}
{\bf Acknowledgments:} 
M. Cveti\v{c} has advocated for some time the idea that black hole 
entropy arises from a single effective string theory comprising 
two chiral sectors. We would like to thank J. Gauntlett, A. Tseytlin, 
and particularly M. Cveti\v{c} for discussions and V. Balasubramanian for 
carefully reading the manuscript. S. Gubser pointed out that the
units are not as stated in the first version of this paper. This work was 
supported by DOE grant DE-AC02-76-ERO-3071.

\end{document}